# Visual Data Mining of Genomic Databases
# by Immersive Graph-Based Exploration

N. FÉREY  
LIMSI/CNRS  
ferey@limsi.fr

P.E. GROS  
LIMSI/CNRS  
gros@limsi.fr

J. HÉRISSON  
LIMSI/CNRS  
herisson@limsi.fr

R. GHERBI  
LIMSI/CNRS  
gherbi@limsi.fr

Université Paris SUD XI  
Bâtiments 508 et 502bis  
91403 ORSAY (France)  
(+33) 1 69 85 81 64

**Abstract**

Biologists are leading current research on genome characterization (sequencing, alignment, transcription), providing a huge quantity of raw data about many genome organisms. Extracting knowledge from this raw data is an important process for biologists, using usually data mining approaches. However, it is difficult to deals with these genomic information using actual bioinformatics data mining tools, because data are heterogeneous, huge in quantity and geographically distributed. In this paper, we present a new approach between data mining and virtual reality visualization, called visual data mining. Indeed Virtual Reality becomes ripe, with efficient display devices and intuitive interaction in an immersive context. Moreover, biologists use to work with 3D representation of their molecules, but in a desktop context. We present a software solution, *Genome3DExplorer*, which addresses the problem of genomic data visualization, of scene management and interaction. This solution is based on a well-adapted graphical and interaction paradigm, where local and global topological characteristics of data are easily visible, on the contrary to traditional genomic database browsers, always focused on the zoom and details level.

**CR Categories:** H.5.1 [Information interfaces and presentation]: Multimedia Information Systems – *Artificial, augmented, and virtual realities.* I.3.7 [Computer Graphics]: Three-Dimensional Graphics and Realism – *Virtual reality.* J.3 [Life and Medical Sciences]: Biology and genetics.

**Keywords:** Virtual Reality, Immersive Exploration, Human-Computer Interaction, Genomic Data, Graph-based Visualization.

## 1. Introduction

The last years witnessed a continued growth of the amount of data being stored in biologic databanks. Often the data sets are becoming so huge, that make them difficult to exploit.

Extracting knowledge from this raw data is an important process for biologists, using usually data mining approaches. However, it is difficult to deals with this genomic information using actual bioinformatics data mining tools, because data becomes very huge in quantity. For example the capacity of DNA microarray data increased by thousand in several years. Even the best bioinformatics visual data mining tools on this kind of data, such as the innovative and famous hierarchical visual clustering of Eisen et al. [1998] do not achieve to deal with this size increasing. The advances in virtual reality and data visualization have thus creating increasing need for graphical tools and techniques to aid in large genomic data analysis. For example, the limit of the desktop context in the Eisen's solution, leaded Kano et al. [2002] to adapt this paradigm into an immersive context. New solutions were developed in order to deal other kind of huge data, such as huge molecule. *ADN-Viewer* [Gherbi. and Hérisson 2002] exploits the advantages of a virtual context with large display, to deals with huge nucleic molecule, and offers biologists a new representation of their huge DNA sequences, by representing its predicted 3D architecture, according to it textual sequence (list of A, C, G, T) and biophysical model. Sharma et al [2002] proposed *Atomsviewer*, a similar solution in an immersive context, in order to explore billion-atom molecules. However, there are other kinds of genomic information relating to genes or molecules, recorded in structured format within many genomic databanks. *Sequence World* [Rojdestvenski et al. 2000] proposes the first solution in an immersive context, in order to explore this kind of huge factual genomic databanks. Nevertheless, and this solution deals only with annotated gene sequence databanks such as *GenBank*, solution, and does not address the problem of heterogeneity.

As *Sequence Word*, this paper presents a visual mining approach, in an immersive context. However, our solution allows biologists to explore and manage huge and heterogeneous genomic data, not only annotated sequence databanks. Our solution is based on a well-adapted graphical and interaction paradigm for genomic data, where global topological characteristics of data are easily visible, on the contrary to traditional genomic database browsers, always focused on the zoom and details level. First, we present in how we address the problem of the format heterogeneity of this kind of databases, in order to explore them with a common visualization paradigm. We explain then how our software deals with huge genomic data, using a specific data representation, an immersive context and simple scene management. Finally, we present some results and experiments produced by *Genome3DExplorer*, leaded by biologists on various sets of biological data.

## 2. Materials and Methods

### 2.1 Database Visualization

Although genomic databases are very heterogeneous in their format or quality, they involve nevertheless some common characteristics. Indeed genomic databases are often focused on biological objects of interest (protein, gene, etc.), described by a set of attributes. Attribute values can be of string type (label, genetic sequence), numerical type (like alignment score), or symbolic type (type of interaction such as positive retroaction). Moreover, these objects are often compared one to another by a measurement in these databanks. (Sequence alignment score, functional similarity…).

In order to store into a common format these various genomic data, taking account of their characteristics, we decide to use in *Genome3DExplorer* a specific graph language. In this graph, nodes stand for focused biological objects of interest of the genomic database, and edges stand for binary relationships between the object into this database. Each node and edge contains a set of values (symbolic or numerical). This language based on multi-valued graph allows us to deals with most commonly genomic databanks, and allows us to use the efficient drawing algorithms coming from the graph theory.

Visualizing these data by graph, suppose that the user choice how he wants to map nodes and edges values to graphical 3D graph characteristics. As shown in Table 1, node values can be map to node size, node transparency, node color, or node shape in the visualization. Edge values can be map to edge color, edge transparency, edge shape, or edge length.

Table 1: Data mapping to 3D graph characteristics

| Object value(Node) | | Relationship value (Edge) | |
|---|---|---|---|
| symbolic | numerical | symbolic | numerical |
|  | position : x,y,z |  |  |
|  | size |  | weight |
|  |  |  | length |
| predefined colors: pink, green, purple… | r, g, b | predefined colors pink, green, purple… | r, g, b |
| predefined shapes sphere, cube |  | predefined shapes line, cylinder… |  |
|  | transparency |  | transparency |

Mapping numerical measurements between two genomic objects of interest, by edge length, is the key point of our approach. Indeed this mapping allows us to position this object between us in 3D space, and translate the data, into a graphic representation, respecting the topology of its data.

However, mapping numerical values to an edge length adds geometrical unsolvable constraints for drawing graph in 3D Euclidian space. For example, Figure 1 shows an undrawable graph in a Euclidian space (3 vertices and 3 edges) respecting the edges lengths.

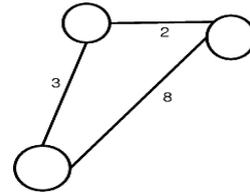

Figure 1: Undrawable graph with edge length constraint

To address this kind of problem, *Genome3Dexplorer* uses method based on the Eades's approach [1984], and improved by Fruchtermann and Reingold [1991], called force-directed placement, which consists in simulating two kinds of force between nodes. In order to place two connected nodes respecting a distance constraint coming from numeric data mapping, Eades proposed to draw graph nodes with a random placement (see Figure 2 (1)), and to apply forces these nodes. In an iterative way, the algorithm compute these forces, in order to minimize the difference between the real edge length connecting two nodes in the graph, and the aimed edge length. These nodes move then according to the forces applied on them. Moreover, repulsion forces are applied on two close nodes, which are not connected, in order to avoid overlap. After several iterations, this dynamic property allows the system to converge into a solution where all the distances are as closed as possible as desired edges lengths.

However, this simulation process is permanently applied on the graph, as shown in Figure 2 on a simple graph. Indeed user can edit by adding or deleting nodes and edges of the graph, and so modify the topology of data; which must be rendered in interactive time.

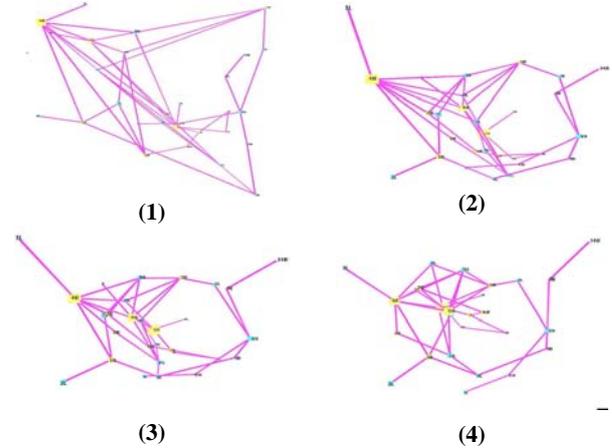

Figure 2: Simulation process at start (1: random placement), after 0.3 sec (2), 0.5 sec (3), 1 sec (4: stabilized graph).

So we had to decrease the high complexity of this basic approach. The repulsion process takes too many time for each iteration (complexity in n2 for each iteration, where n is the number of nodes), and we decided to give up the repulsion forces. In fact, repulsion was very useful in a 2D graph-drawing context, but not really in a 3D space, where a tree drawing instead of a random one is efficient to avoid overlap. For each iteration, we obtain a complexity proportional to n + e, where n is the number of nodes and e the number of edges. This complexity allows us to apply force simulation to a graph of some tens of thousands of nodes and edges in interactive time. For bigger graph, the simulation must be processed offline before displaying, and then we just apply force simulation on the nodes into the cone view.

## 2.2 Immersive Exploration

In order to visualize huge quantity of data, we choice to use the advantages of large surface displaying offered by a virtual reality context. Moreover, the building and management of the genomic database is a strongly collaborative work and the immersive context is well adapted for this kind of job. Classical human-machine interaction paradigms are well adapted for desktop environment. However, they are rapidly useless when 3D objects are manipulated in 3D space. The stereoscopy mechanism and large visualization area make it possible to decrease significantly the gap between the virtual space and the real one. For these reasons, *Genome3DExplorer* was integrated within the LIMSI Virtual Reality platform. This immersive environment platform is equipped with two components: hardware part and software architecture. For the hardware one, we use two rear-projected orthogonal screens (2mx2m size) and a graphic cluster of four pc. For the software part, on the one hand, active stereovision and graphic clustering with the pc cluster and the two screens are managed thanks to *Chromium* [Humphreys et al. 2002]. On the other hand, a middleware developed at LIMSI-CNRS, *VEServer* [Touraine et al. 2002], manage various 3D and immersive devices as data glove for gesture recognition, speech recognition system, 3D position and orientation tracking system, immersive mouse.

## 2.3 Immersive Interactions

The genomic database exploration and perception are significantly increased using more intuitive immersive devices. This allows the user to focus completely on his task.

### 2.3.1 Classical immersive devices

*Genome3DExplorer* uses two alternative 3D pointing devices: the WANDA$^{TM}$ or the data glove. It has proved very useful to draw a virtual ray having for origin the WANDA$^{TM}$ or the glove and for direction the orientation of the wand or the finger. We can thus easily point or select any object while navigating, in order to offer information feedback to the user.

For other interaction types, we use speech recognition to add graph node, to add new edge between two selected graph nodes, or to delete selected node or edge. We use this interaction modality to stop, start or modify the force-directed placement algorithm (repulsion or attraction factor). Finally, we use speech recognition to hide or show node, edge, or label of the graph.

### 2.3.2 Spaghetti interaction request

Taking account the force simulation is permanently applied on the graph, when we catch and move a selected graph node, all the nodes directly or indirectly connected to it, follow this moving. We can see this interaction as an extraction vicinity request for a specified data node. This interaction is yet a prototype, but appears as very intuitive for the users; taking into account the context of interactive force-directed placement algorithm applied permanently on the graph.

## 3. Results and Experiments

We present in this section three experimentation example of our system leaded by biologist users on different kind of genomic data.

## 3.1 Yeast Gene Block Duplications

This system was first used in order to visualize gene duplications in the yeast chromosomes with a synthetic way. In this experiment, each object is one of the yeast chromosome arms. For each object, the values are chromosome name, chromosome size, chromosome side (right or left arm). In each relationship between chromosomes, the values are the number of same gene shared by two chromosomes. Compared to the detailed 2D representation of chromosome gene duplication in the Figure 3, we can directly see in Figure 4 that several chromosome arms, like 4R, are in the centre of the global gene duplications, whereas other chromosome arms take placed in periphery. This representation helps biologist to launch a work on correlation between chromosome placement in the cells and the gene duplication between Yeast chromosomes during evolution. This first experiment does not show the advantage of immersive 3D graph because data are simple, but highlights the capabilities of this representation to offer a more synthetic visualization of his data sometimes more useful for biologist.

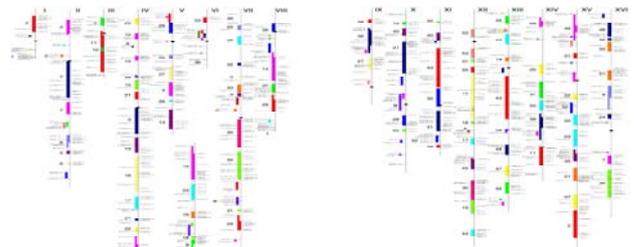
Figure 3: Traditional 2D visualization (16 Yeast chromosomes)

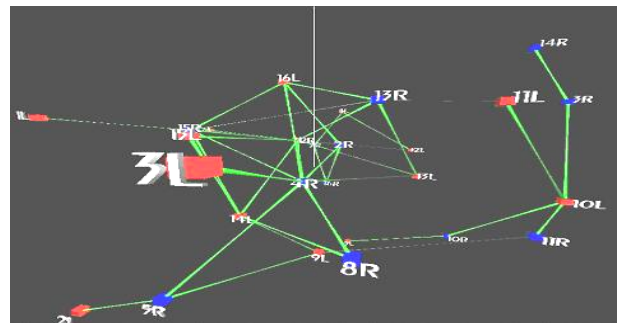
Figure 4: Gene duplications in a Yeast chromosome

## 3.2 Chromosome Architecture Using Co-Regulation Criteria

The aim of this second experiment was to visualize the 3D architecture of the chromosome 3 of Yeast, according to co-regulation consideration. Indeed, in the cells, two genes may work at the same time, and biologists call these genes "co-regulated genes". We can see in, the chromosome sequence in blue, the gene along this sequence representing by sphere, and the co-regulation constraints between these genes representing by red edges in the center. As in the first experiments, data are not huge, but some of red edges (co-regulation data) were added during the interactive computation (see Figure 5) of force-directed placement. This function allows biologist to see the consequence of adding new edge or node, (corresponding two new biological data), on the 3D graph topology in real time.

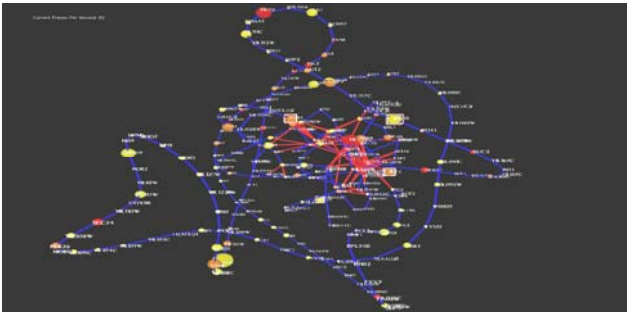
Figure 5: Gene co-regulations data in a Yeast chromosome 9

### 3.3 Exploring Microarray Data

In this experiment, biologist wants to explore huge DNA microarray data. A DNA microarray is a set of gene behaviors according to different experimental conditions. This behavior is called profile. When two genes have a same or opposite profile, it could mean these genes work in the same biological process. This is very important for biologists to identify all of genes, which take part of a biological process, in order to understand the how this process works. In this experiment, biologist choices to represent gene profiles by green cubes. Correlations between two profiles are represented by blue edges and opposite correlations and red edges. Biologists validated our exploration method on his huge data, because many genes in the same sub-network are known as working into the same biological process. Our method allows biologist to focus on unknown gene into known genes cluster.

As opposed to the precedent experiments, we can see in the Figure 6 the limit of the 2D graph representation because of the huge quantity of data. Active exploration in an immersive context is necessary in order to understand this kind of data. Using this immersive context is a strong point to explore and manage huge quantity of data.

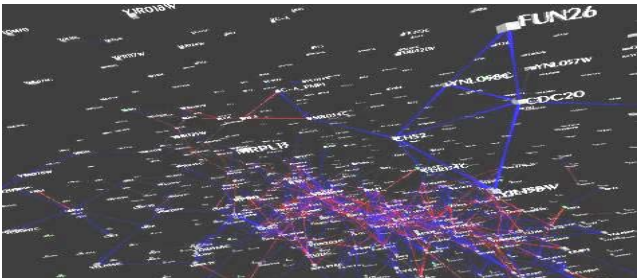
Figure 6 : Yeast correlation expression profiles network

### 4. Conclusion

The immersive aspect and the possibility of exploring large quantities of data in a synthetic way, constitute the strong points of this system, because it gives a global vision of the structure under unclaimed of the data. These characteristics are particularly interesting when biologists wish to explore a mass of data without precisely knowing what they seek. This visualization allows users a more direct interpretation of results coming from large calculations. Our solution is a new way to explore and analyze various sets of genomic databanks, because local and global topological characteristics of data are not easily visible using traditional genomic database browsers, always focused on the zoom and details level, but not at the overview level.

### Acknowledgements

Our thanks to VENISE team (*Virtual ENvironment for Immersive Simulation and Experiments*: www.limsi.fr/venise) for many technical advices and for providing VR devices and software support. We thank also the PPF Bioinformatics project of Paris-Sud University for grant support. We thanks too all biologists for testing our system and for providing us usability feedback, particularly Oriane Matte-Tailliez, and Claire Toffano-Nioche.